\PassOptionsToPackage{bookmarks=false}{hyperref}
\documentclass[sigconf]{acmart}
\renewcommand\footnotetextcopyrightpermission[1]{} %
\AtBeginDocument{%
	\providecommand\BibTeX{{%
	\normalfont B\kern-0.5em{\scshape i\kern-0.25em b}\kern-0.8em\TeX}}
}

\copyrightyear{2020} 
\acmYear{2020} 
\setcopyright{acmcopyright}\acmConference[SEAMS '20]{IEEE/ACM 15th International Symposium on Software Engineering for Adaptive and Self-Managing Systems}{October 7--8, 2020}{Seoul, Republic of Korea}
\acmBooktitle{IEEE/ACM 15th International Symposium on Software Engineering for Adaptive and Self-Managing Systems (SEAMS '20), October 7--8, 2020, Seoul, Republic of Korea}
\acmPrice{15.00}
\acmDOI{10.1145/3387939.3388614}
\acmISBN{978-1-4503-7962-5/20/05}

\usepackage{preamble}

\begin{document}
\title{Towards Highly Scalable Runtime Models with History}
\author{Lucas Sakizloglou,
Sona Ghahremani,
Thomas Brand,
Matthias Barkowsky, and
Holger Giese}
\email{<first-name>.<last-name>@hpi.de}
\affiliation{%
  \institution{Hasso Plattner Institute, University of Potsdam, Germany}
}
\renewcommand{\shortauthors}{L. Sakizloglou et al.}

\begin{abstract}
Advanced systems such as IoT comprise many heterogeneous, interconnected, and autonomous entities operating in often highly dynamic environments. Due to their large scale and complexity, large volumes of monitoring data are generated and need to be stored, retrieved, and mined in a time- and resource-efficient manner.  Architectural self-adaptation automates the control, orchestration, and operation of such systems. This can only be achieved via sophisticated decision-making schemes supported by monitoring data that fully captures the system behavior and its history.

Employing model-driven engineering techniques we propose a highly scalable, history-aware approach to store and retrieve monitoring data in form of enriched runtime models. We take advantage of rule-based adaptation where change events in the system trigger adaptation rules. We first present a scheme to incrementally check model queries in the form of temporal logic formulas which represent the conditions of adaptation rules against a runtime model with history. Then we enhance the model to retain only information that is temporally relevant to the queries, therefore reducing the accumulation of information to a required minimum. Finally, we demonstrate the feasibility and scalability of our approach via experiments on a simulated smart healthcare system employing a real-world medical guideline.

\end{abstract}
\maketitle

\section{Introduction}\label{sec:intro}
Advanced systems such as IoT comprise many interconnected, heterogeneous, and autonomous entities operating in often highly dynamic environments~\cite{DBLP:journals/iotj/CatarinucciDMPP15}. Such systems are challenging to operate:
On one hand, complexity of the underlying architecture and heterogeneity of the involved entities require sophisticated orchestration schemes~\cite{MANOGARAN2018375}---architectural self-adaptation~\cite{Garlan+2009} has been shown to be competent in tackling this challenge via efficiently managing large dynamic architectures at runtime~\cite{Ghahremani.2017.Efficient}.
On the other hand, due to their large scale, these systems generate large volumes of data demanding to be stored, retrieved, and mined in a time- and resource-efficient manner~\cite{DBLP:journals/access/BakerXA17}---technologies such as cloud computing and big data analytics have emerged to improve computational capabilities and facilitate efficient data storage capacities~\cite{SAKR201644}.

In the special case of healthcare systems, big data analytics services~\cite{Coronato2010} can monitor and detect various measurements such as patient vitals which can be provided to clinicians to support them in diagnosis. Recent advancements in healthcare systems as IoT are focused on automating the diagnosis phase to minimize or remove the need to involve a clinician for routine procedures. However, the proposed methodologies often do not consider the resource constraints such as storage and computational power to be a concern -- although relevant in a self-adaptation scenario -- and presume trivial resource availability~\cite{MANOGARAN2018375}.

Employing self-adaptation for automating the control and operation of large IoT systems requires resource-aware schemes in the adaptation loop. 
For architectural self-adaptation, a runtime model is usually considered~\cite{Garlan+2009}. The runtime model reflects a snapshot of the system architecture at a given time and, in order to enable resource-awareness, is enriched with information necessary to analyze and plan the adaptation. 
However, certain systems require knowledge about the evolution history and prior interactions to allow for sound decision-making in the adaptation loop, e.g. for self-explanation \cite{DBLP:conf/iceccs/BencomoWSW12} or for the detection of persistent adaptation issues \cite{Ghahremani.2017.Efficient}. Moreover, adaptation based on temporal information might be integral to the application domain, as is the case for healthcare where temporal requirements are key~ \cite{Combi_2012_ModellingTemporalDatacentricMedicalProcesses}. Besides resource-awareness, adaptation loops for such systems require history-awareness. Snapshot-based representations and related querying schemes lack the careful treatment required for preserving and respectively retrieving such temporal information.

In this paper we propose a highly scalable, history-aware approach for runtime models. Our approach benefits from employing model-driven engineering techniques, where the system layout and monitoring data is captured by runtime models causally connected to the underlying system. It also relies on rule-based architectural adaptation, where the changes, i.e., adaptation issues, trigger adaptation rules. The runtime model captures the evolution history and serves as a unique point of reference, independent from the specific model queries which are subject to change. This promotes separation of concerns. 

Our contributions are as follows. We employ model queries on the runtime model to identify adaptation issues and a graph-based temporal logic to specify temporal requirements for queries. We extend our work in~\cite{Giese_2019_MetricTemporalGraphLogicoverTypedAttributedGraphs} by presenting a scheme to incrementally check the queries against a runtime model with history. We then enhance the runtime model to retain only information that is temporally relevant to the queries and therefore reduce the perpetual accumulation of historical information to a required minimum. Finally, we demonstrate the feasibility and scalability of the approach by evaluating it on a simulator of a smart healthcare system~(SHS) and a real-world medical guideline.

The rest of the paper is organized as follows.
The foundations and required building blocks for the proposed approach are discussed in \autoref{sec:prereq}.
 In \autoref{sec:approach} we introduce the proposed approach for obtaining and mining history-aware runtime models in a  scalable and memory-efficient manner.
 \autoref{sec:eval} presents an implementation and evaluation of the approach on a simulated SHS.
 We discuss the related work in \autoref{sec:related} and \autoref{sec:conclusion} concludes the paper with an outlook on planned future work.

\section{Foundations}
\label{sec:prereq}

\subsection{Architectural Self-adaptation and Runtime Models}\label{subsec:SASwRTM}
To realize self-adaptation, a software system is equipped with a \textit{MAPE-K} feedback loop that \underline{m}onitors and \underline{a}nalyzes the system and, if needed, \underline{p}lans and \underline{e}xecutes an adaptation of the system, which is all based on \underline{k}nowledge~\cite{Kephart&Chess2003}.
Self-adaptation can be generally achieved by adding and removing components as well as connectors among components in the system architecture~\cite{MageeKramer1996}, therefore, many researchers consider the \textit{software architecture} as an appropriate abstraction level (e.g.,~\citeN{Oreizy+1999,Garlan+2009}). %
For this purpose, the feedback loop maintains a \textit{runtime model} as part of its knowledge to represent the architecture of the system, which is \textit{causally connected} to the system, that is, any relevant change of the system is reflected in the model, and vice versa~\cite{Blair+2009}. Thus, the phases of the MAPE-K feedback loop operate on the runtime model to perform self-adaptation.
Self-adaptation is realized via (self-)\emph{adaptation rules}. We focus on adaptation rules with an \emph{event-condition-action} form. In detail: Changes of the systems are captured as events; Occurrences of change events influence whether a condition is satisfied and hence if an event-condition-action rule is applicable; An action defines the required system changes for adaptation. Such rules capture both analysis (evaluating the condition) and planning (applying the action) activities. In this paper, we are interested in efficient evaluation of the conditions during the analysis phase.

\vspace{-2mm}
\subsection{Graph-based Models and Rules}

Our approach adopts the established practice of representing a runtime model as a graph, where architectural components are modeled as vertices, connectors between components as edges, and further information about the components are encoded in attributes~\cite{VG10}. Employing a graph-based runtime model as representation of the underlying system allows for model-driven engineering (MDE) techniques to be employed within the MAPE-K phases \cite{France+Rumpe2007}, where such a graph conforms to a language for runtime models, specified by means of a \emph{metamodel}. The metamodel defines types of vertices, edges, and attributes and restricts the structure of valid model instances. Formally, these concepts rely on \emph{typed, attributed graphs} where graphs are typed over a \emph{type graph} \cite{DBLP:conf/gg/EhrigPT04}.

Employing MDE, in our earlier work~\cite{Vogel+2009,VogelNHGB10} we showed how model queries based on graph patterns can be used during the analysis phase to identify adaptation issues in the runtime model, whereas the action that describes how to adapt the model to resolve these issues is applied by in-place model transformations based on \emph{typed, attributed graph transformation} \cite{DBLP:conf/gg/EhrigPT04}.
In summary (and similar to the scheme presented in~\cite{Ghahremani.2017.Efficient, SG-TAAS}), let $G$ be a graph representation of the runtime model, effectively the system architecture. A structural fragment of the architecture can be described as a pattern $P$ of a set of patterns $\mathcal{P}$.
Then, an adaptation rule $r$ can be characterized by a left-hand side pattern $LHS$ and a right-hand side pattern $RHS$, which define the precondition and postcondition of an application of $r$, respectively. A match $m$ of $LHS$ in $G$ corresponds to an occurrence of the pattern $LHS$ in the model $G$ and, intuitively, identifies a part of the runtime model where the adaptation should occur in the architecture. For the rest of the paper, we use the terms \emph{query} and \emph{pattern} interchangeably.

In certain cases, simple graph patterns are not sufficient as a language 
for specifying application conditions of adaptation rules, for instance if the existence of certain model elements should be prohibited. The established logic of \emph{Nested Graph Conditions} (NGC)~\cite{Habel_2009_Correctnessofhighleveltransformationsystemsrelativetonestedconditions} can be employed to equip the $LHS$  pattern of an adaptation rule with constraints which are expressively equivalent to first-order logic on graphs~\cite{Courcelle_1997_TheExpressionofGraphPropertiesandGraphTransformationsinMonadicSecondOrderLogic}. It uses standard logic operators, i.e., negation ($\neg$), existential quantification ($\exists$), conjunction ($\wedge$),  implication ($\rightarrow$), universal quantification ($\forall$), and nesting of patterns to bind graph elements in outer conditions and relate them to inner (nested) conditions, thereby allowing for more elaborate specifications. 

The search for matches of patterns (potentially restricted by constraints, as in NGC) in a given graph $G$ is called \emph{pattern matching}. Depending on the size of the graph and the pattern, pattern matching can be considerably complex and time-consuming. Techniques such as \emph{local search}, where the search space is reduced by starting the search from a specific element of the graph, have been employed to reduce the required effort~\cite{Hildebrandt14}.

\subsection{Temporal Requirements}\label{subsec:temporal-conditions}
Besides the description of a structural fragment of the architecture, in this work we are also interested in describing the evolution of such a fragment over time. In order to achieve this, a rule specification language should enable the definition of \emph{temporal requirements}, that is, expectations on the change of a structure over time, which is not inherently supported by NGC. For this purpose, 
we rely on our previous work in \cite{Giese_2019_MetricTemporalGraphLogicoverTypedAttributedGraphs} to specify application conditions for rules in \emph{Metric Temporal Graph Logic} (\MTGL).
The logic builds on NGC (supports all NGC connectives) and the \emph{Metric Temporal Logic} (\MTL) \cite{Koymans_1990_Specifyingrealtimepropertieswithmetrictemporallogic} to introduce the temporal operators \emph{for-all-new} (\MTGCforallNewSimple{}), \emph{until} ($U_I$, where $I$ is an interval over $\rm I\!R_{0}^{+}$), and \emph{eventually} ($\lozenge_I$) that enforce restrictions on patterns based on temporal requirements for the future. Based on findings that show past operators do not affect the expressiveness of the logic \cite{Havelund_2018_RuntimeVerificationFromPropositionaltoFirstOrderTemporalLogic} and how \MTL formulas with future operators can be translated into formulas with past operators \cite{Hunter_2013_ExpressiveCompletenessforMetricTemporalLogic}, \MTGL is in capacity to express requirements also about the past, i.e., the past temporal operators \emph{since} ($S_I$) and \emph{once} $(\blacklozenge_I)$. Temporal operators express temporal requirements over matched patterns such as: ``\emph{for all matches of the pattern $P$ that are observed for the first time in $G$}'' (\MTGCforallNewSimple{P})\footnote{For simplicity, the formulation omits certain syntactic elements of \MTGL.} and ``\emph{the pattern $P$ must once have existed in the graph  $G$ in the last 10 time units}'' $(\blacklozenge_{[0,10]}\,\exists \,P)$.\footnote{The interpretation of a time unit depends on the system's time-tracking granularity.}

\subsection{Example: Smart Healthcare System}\label{subsec:running-example}
As a running example, we employ a simulated Smart Healthcare System (SHS) where sensors periodically collect physiological measurements of patients, i.e., data such as temperature, heartbeat rate, or blood pressure. 
Certain medical procedures are automated and performed by devices, such as a smart pump administering medicine based on the collected patient measurements---as otherwise a clinician would be doing. 
\autoref{fig:approach-metamodel} presents an excerpt of the metamodel of the envisaged SHS---modeling features that enable history-awareness are explained later in \autoref{subsec:RTMwHistory}.
Every change in our simulated SHS, i.e. a new sensor measurement or an activated pump, generates an \emph{event}. These events are (transparently) processed and reflected on the runtime model: For instance, during monitoring phase, when a pump is activated, a node of type \emph{Pump} is added to the model; when a \CLASSNAME{Pump} takes a reaction or a \CLASSNAME{PatientSensor} emits a new datum, a \CLASSNAME{StringValue} node is generated with the value of the reaction  respectively datum---in the following, we refer to these nodes in short as \emph{reaction} and \emph{datum}.

\begingroup
\setlength{\textheight}{\the\dimexpr19.3cm+1\baselineskip}
\setlength{\textfloatsep}{.7\baselineskip plus 0.2\baselineskip minus 0.5\baselineskip}
\setlength{\belowcaptionskip}{0pt}
\setlength{\abovecaptionskip}{0pt}
\begin{figure}
		\vspace{-3mm}
	\centering
	\includegraphics[width=\linewidth]{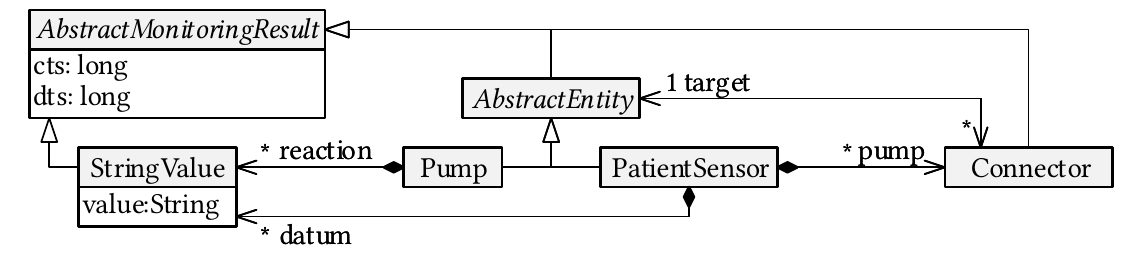}
	\caption{\label{fig:approach-metamodel}
		Excerpt of SHS Metamodel.}
	\vspace{-3mm}
\end{figure}
\endgroup

Based on the employed SHS, we attempt to enact an instruction from a  medical guideline on the prevention of infective endocarditis for patients who undergo a dental procedure~\cite{Wilson_2007_PreventionofInfectiveEndocarditisAGuidelineFromtheAmericanHeartAssociation}.
The (simplified) instruction reads: ``An antibiotic should be administered within one hour before the procedure. If antibiotic is inadvertently not administered before, it may be administered up to 2 hours after the procedure''.
Based on the instruction, we extract \emph{event-condition-action} rules (cf. \autoref{subsec:SASwRTM}), i.e., as a result of system changes captured in form of events (during monitoring), if a condition holds (during analysis) then an action should be taken (during planning).

The instruction contains two statements which both assume a procedure has taken place: The first concerns the antibiotic administration within one hour before the procedure and the second the alternative action in case administration has been omitted. In the following, we focus on the latter---since its condition accounts also for the former. We extract the rule $\mathcal{C}$: \emph{If a patient has undergone a procedure and an antibiotic has not been administered up until 1 hour before the procedure} (condition)\emph{, it may be administered up to 2 hours after the procedure} (action).
In this work we focus on the analysis phase of the adaptation loop, i.e., detecting when a condition holds and, as a result, an adaptation of the system is required. 
If the condition holds the planning phase would enact the action.

In our simulation, instruction tasks are represented as follows: If a patient has undergone a procedure a \CLASSNAME{StringValue} node is generated by a \CLASSNAME{PatientSensor} with the value \emph{op}. If a \CLASSNAME{Pump} has administered antibiotics, a \CLASSNAME{StringValue} is generated with the value \emph{anti}.
\section{The \APPROACHNAME Querying Scheme}\label{sec:approach}

In the following we describe the proposed scheme for checking queries with temporal requirements against runtime models with history in an online, scalable, and sustainable manner. We name the scheme \APPROACHNAME---from its key features: the \emph{In}cremental checking and the support for \emph{Tempo}ral requirements.
The core of \APPROACHNAME consists of two steps. First, in step $\sigma_1$ (\autoref{subsec:RTMwHistory}), the metamodel of the runtime model is modified so that each element has a creation and deletion timestamp which are used to capture temporal information about every change in the system. 
The second step $\sigma_2$ (\autoref{subsec:incremental-temporal-matching}) entails the translation of a temporal model query such as $\mathcal{C}$ from our running example, into a structural query with application conditions that capture the temporal requirements and can be incrementally checked.
The pruning step $\sigma_3$ (\autoref{subsec:sustainable-memory}) concerns the derivation of pruning rules based on the temporal requirements of the queries. Pruning rules can be optionally employed to prune the runtime model with history so that elements that are not required by queries are not retained.
\subsection{Enriching the Runtime Model with History}\label{subsec:RTMwHistory}
\begingroup
\setlength{\textheight}{\the\dimexpr19.3cm+1\baselineskip}
\setlength{\textfloatsep}{.7\baselineskip plus 0.2\baselineskip minus 0.5\baselineskip}
\setlength{\belowcaptionskip}{0pt}
\setlength{\abovecaptionskip}{0pt}
\begin{figure}
	\centering
	\vspace{-3mm}
	\includegraphics[width=\linewidth]{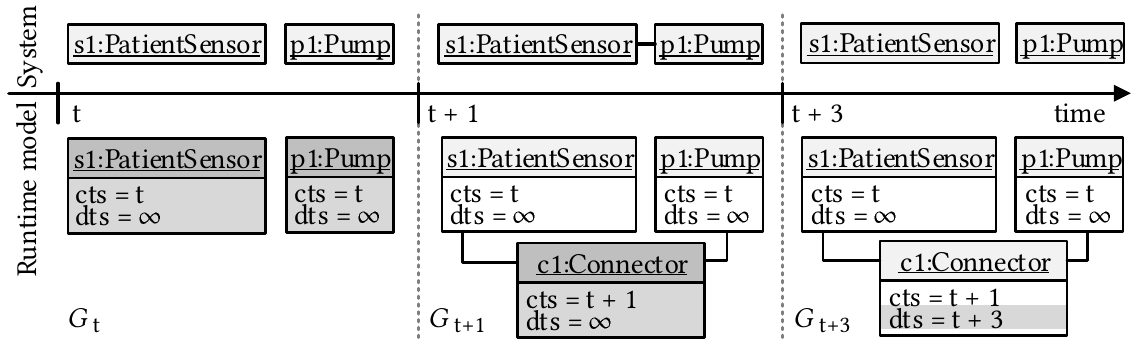}
	\caption{\label{fig:approach-evolving_model}Capturing System History in the Runtime Model.}
	\vspace{-3mm}
\end{figure}
\endgroup	
\vspace{-1mm}
Step $\sigma_1$ enriches the metamodel of the runtime model to enable history-awareness (see \autoref{fig:approach-metamodel}). We refer to instances of history-aware metamodels as \emph{runtime models with history}. In such a metamodel, every change, i.e., update, creation, or deletion of an element of the system, is a result of a monitoring activity. Hence, all  nodes of the metamodel inherit from  \CLASSNAME{AbstractMonitoringResult} which uses two timestamps to mark the lifespan of an element (\emph{cts} for creation and \emph{dts} for deletion). Physical components of the SHS, such as the patient sensor (\CLASSNAME{PatientSensor} node) and the pump (\CLASSNAME{Pump} node) inherit from the class \CLASSNAME{AbstractEntity}. The \CLASSNAME{StringValue} node captures sensor measurements and pump reactions, both a string data type. Connections between components are captured by the \CLASSNAME{Connector}. The nodes \CLASSNAME{StringValue} and \CLASSNAME{Connector} as well as the \emph{cts} and \emph{dts} timestamps are domain-independent features required to capture the history of values or connections in the model.

Assuming a sequence of timestamped change events generated by the system at the point in time when they occur (hereby mentioned as \emph{timepoint}), each change is applied to the runtime model by identifying or creating the corresponding model elements and setting the values of the \emph{cts} (or, if applicable, the \emph{dts}) timestamp. If the event entails a creation, the \emph{dts} is set to $\infty$ by default (in practice, the maximum value of the long data type). For instance, as shown in \autoref{fig:approach-evolving_model}, when a sensor \emph{s1} and a pump \emph{p1}  is activated in the system, the runtime model reflects this change by creating two nodes \emph{s1} and \emph{p1} and setting the \emph{cts} value to the timestamp $t$ of the creation (graph $G_t$) and the \emph{dts} to $\infty$. When later, at a timepoint $t+1$, \emph{p1} is connected to \emph{s1}, this connection is reflected in $G_{t+1}$ by the creation of a \emph{Connection} node \emph{c1} with a \emph{cts} equal to $t$+1. A disconnection at timepoint $t$+3 is reflected in $G_{t+3}$ by the node's \emph{dts} being set to $t$+3, that is, the element is not actually removed from the model.
\subsection{From Temporal to Structural Conditions}
\label{subsec:incremental-temporal-matching}

In the following we present the inner workings of the step $\sigma_2$ of \APPROACHNAME. Effectively, $\sigma_2$ entails the translation of patterns with both structural and temporal application conditions into patterns with only structural application conditions and attribute constraints, following a similar operation in \cite{Giese_2019_MetricTemporalGraphLogicoverTypedAttributedGraphs}. 
Temporal application conditions are encoded in the translated pattern by means of constraints over the \emph{cts} and \emph{dts} attributes of the related nodes. The translation enables the usage of highly efficient techniques for pattern matching, thereby expediting the detection of LHS of rules and improving the overall performance of the adaptation via reducing the execution time of the adaptation loop. 
Throughout the section we provide examples based on the instruction $\mathcal{C}$ from the medical guideline and the SHS (\autoref{fig:approach-metamodel}).

Let $\phi_1$ and $\phi_2$ be the structural patterns representing the two statements in the condition of $\mathcal{C}$  (``If a patient has undergone a procedure''  and ``an antibiotic has not been administered up until 1 hour before the procedure'' respectively). In the runtime model of our simulated SHS, $\phi_1$ reflects that a new \emph{datum} (\emph{StringValue} node) has been emitted (from a sensor) with the value \emph{op}. Similarly, $\phi_2$ reflects that a new \emph{reaction} has been created (from a pump) with the value \emph{anti}. Based on the temporal requirements, let the formula $\phi$ for the condition of $\mathcal{C}$ in \MTGL be $\phi = \MTGCforallNewSimple{\phi_1} \rightarrow \blacklozenge_{[0,3600]} \exists  \, \phi_2$, which intuitively states that every time a new match of $\phi_1$ is found in $G$, $\phi_2$ must have once existed in $G$ in the previous hour (and assumes that the system uses seconds to generate timestamps).

Except for the structural patterns, the formula $\phi$ contains two types of operators. On the one hand, those that express structural application conditions and are formulated by NGC, e.g., existential quantification for asserting the pattern's existence.
On the other hand, those that express temporal ($\blacklozenge$) or both temporal and structural (\MTGCforallNewSimple{}) application conditions and are formulated by \MTGL.
Recall that the objective of this step is to translate $\phi$ into an NGC formula with attribute constraints. We translate the \MTGL temporal application conditions into attribute constraints based on the timestamps of each element and the timepoint $x$, where $x$ is the current timepoint in the system. 

Let $\psi$ be the translated formula, $\kappa$ a set of attribute constraints, and \emph{sat} a boolean function that checks the satisfaction of $\kappa$. 
The operator \MTGCforallNewSimple{} states that only patterns that are new with respect to $x$, that is, found for the first time, are matched. We translate this by asserting the existence of $\phi_1$ ($\exists \,\phi_1$) and adding the attribute constraint $\alpha.$\emph{cts}$\,\,=x$ to $\kappa$, that at least one element $\alpha$ in $\phi_1$ needs to satisfy. 
The operator $\blacklozenge_{[0,3600]}$ states that the pattern $\phi_2$ should have once existed in the previous hour. We translate this by asserting the existence of $\phi_2$ ($\exists\, \phi_2$) and adding the following attribute constraint in $\kappa$ that each element $\beta$ in the pattern should satisfy: $(\beta.$\emph{cts} $< x-$0 $\wedge$ $\beta.$\emph{dts} $> x-$3600), that is, the creation of $\beta$ has occurred at some timepoint before $x$ and its deletion has occurred at some timepoint after $x-$3600. 
The resulting formula is $\psi = (\exists \, \phi_1 \rightarrow \exists \,\phi_2) \, \wedge $ \emph{sat}$(\kappa)$
which intuitively states that both patterns $\phi_1$ and $\phi_2$ must exist in the runtime model with history and that the temporal requirements that refer to 
timestamps of the elements in the patterns and are captured by the attribute constraints in $\kappa$ must be satisfied.

By using only NGC and attribute constraints, we have derived a formula that captures temporal requirements and, on a runtime model with history, allows for efficient querying.

\vspace{-1mm}
\subsection{Sustaining Memory Consumption}
\label{subsec:sustainable-memory}

\APPROACHNAME relies on capturing historical information about the lifespans of events in the runtime model with history. The accumulation of events over time causes the graph to grow in size; model elements whose lifespan is over, such as a connection between two components, are never actually removed from the model, rather the end of their lifespan is marked by an update of their \emph{dts}. 

Based on application domain guidelines, a process similar to a garbage-collector could potentially remove obsolete or unusable elements from the model. For instance, in our running example, a guideline from the SHS administration can specify that all pump reactions, i.e., \emph{reaction} nodes, that are older than a certain amount of time, e.g. a month, can be safely discarded.
For the targeted type of systems however, i.e., IoT governed via self-adaptation, the interim accumulation of data until such a generic guideline is applied could still require an impractical amount of memory for the model (which is typically stored in-memory). 
Moreover, the growing size of the model may lead to deteriorating performance of the pattern matching as more elements have to be considered during the search.
A more refined guideline could take into account elements that are considered in patterns with temporal requirements that refer to a specific (past or future) time window and thus become obsolete when this window elapses. These elements can never be part of a new match and can therefore be removed from the runtime model immediately after the end of their respective time window.

The pruning step $\sigma_3$ of \APPROACHNAME entails the analysis of the formula to calculate such a time window for which elements must be retained. Once the analysis is done, \emph{pruning rules} can be derived. Pruning rules can be used to prune the model by removing obsolete model elements. For instance,  for the formula $\phi$, let $t$ be the current system timepoint: Any \emph{datum} whose \emph{dts} is smaller than $t$ may be removed from the model (together with its edge) as it is not going to be part of a new match---provided no new sensors are added; Similarly, a \emph{reaction} may be removed when its \emph{dts} is smaller than $t$+3600. 
Pruning rules are uniquely defined and can also be systematically derived---even at runtime, if necessary.
Due to time required for searching and removing the elements, pruning incurs an increase in the overall adaptation time. Provided such an increase is acceptable, this step renders the memory consumption of \APPROACHNAME sustainable even in the absence of any external memory-saving measures.
\section{Implementation and Results}\label{sec:eval}
In this section we present an implementation of \APPROACHNAME, together with the employed setting for evaluation experiments, obtained results, and threats to validity. 
\subsection{Implementation}\label{subsec:implementation}

\begingroup
\setlength{\textheight}{\the\dimexpr19.3cm+1\baselineskip}
\setlength{\textfloatsep}{.7\baselineskip plus 0.2\baselineskip minus 0.5\baselineskip}
\setlength{\belowcaptionskip}{0pt}
\setlength{\abovecaptionskip}{0pt}
\begin{figure}
	\centering
	\includegraphics[width=0.85\linewidth]{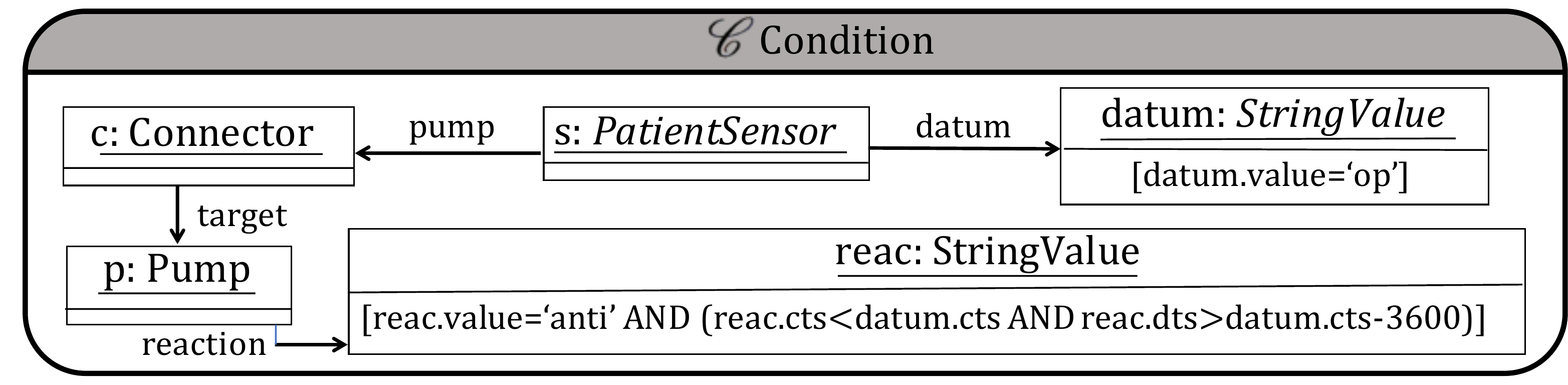}
	\caption{\label{fig:property} Story Diagram for Condition of Instruction $\mathcal{C}$.}
	\vspace{-3.5mm}
\end{figure}
\endgroup

To demonstrate the advantages of \APPROACHNAME, we prototypically implemented the three steps presented in \autoref{sec:approach} for the condition of instruction $\mathcal{C}$ from the running example (see Section~\ref{subsec:running-example}).
Our implementation is based on the Eclipse Modeling Framework (EMF), as the former is a widespread MDE technology for creating metamodels. In the following, we refer to specific EMF features when we describe details of our implementation.

Step $\sigma_1$ is implemented by modeling the metamodel in EMF (cf.~\autoref{fig:approach-metamodel}).
For step $\sigma_{2}$, first we obtain the translated formula $\psi$ featuring patterns with NGC application conditions and attribute constraints as explained in \autoref{subsec:incremental-temporal-matching}. We then map $\psi$ to \emph{Story Diagrams} (SD) \cite{DBLP:conf/tagt/FischerNTZ98}. SD are a visual language supported by a tool based on EMF that allows for the specification of pattern matching tasks. The tasks are executed by means of an interpreter that enables incremental processing of change events via local search~\cite{DBLP:journals/eceasst/GieseHS09}. 
SD can encode NGC, controls the flow of the pattern matching depending on whether a (sub)pattern was matched, and supports the definition of attribute constraints on matched elements. 

The SD for our translated formula $\psi$ is shown in \autoref{fig:property}. 
It starts by searching for a \emph{PatientSensor} that has emitted a \emph{datum} with the value \emph{op}---technically, one that satisfies the constraint on the \emph{value} attribute---which is the first statement of the condition of  $\mathcal{C}$ from our running example ($\exists \, \phi_1$ in $\psi$). Note that the type names of elements that can serve as starting points for the search are italicized. The search proceeds only if this structure has been matched, which captures the implication in $\psi$. If the search proceeds, it searches for a \emph{Pump} connected to the patient (via a \emph{Connector}) and a \emph{reaction} with the value \emph{anti} ($\exists \, \phi_2$ in $\psi$).

Regarding the satisfaction of temporal requirements in $\psi$, i.e.,  \emph{sat}$(\kappa)$, the constraints in $\kappa$ from the operator \MTGCforallNewSimple{} are captured implicitly by triggering an execution of SD only upon the creation of elements of interest, in this case a \emph{datum} or a \emph{PatientSensor} which is enabled by the EMF notification mechanism.
This allows for incremental checking, i.e., upon every relevant change event in the model, we only check for newly found matches. Moreover, searching for matches upon a change event is performed by considering only the recently added element(s) and relevant  adjacent parts, thus taking advantage of local search heuristics which can significantly reduce the pattern matching effort \cite{Hildebrandt14}.

The constraints in $\kappa$ from the operator $\blacklozenge$ are added directly to the pattern elements using OCL constraints \cite{DBLP:conf/tools/KleppeW00a}. 
For clarity, in our implementation we only check the temporal requirements for the \emph{reaction}. Thus, similar to the medical guideline \cite{Wilson_2007_PreventionofInfectiveEndocarditisAGuidelineFromtheAmericanHeartAssociation}, we assume that if a patient has undergone a procedure then a pump and a sensor has been connected to that patient and that the lifespans of pattern elements overlap---in fact, our approach checks when such implicit assumptions do not hold, e.g. in this case, if matching would fail because a pump is not connected. Furthermore, based on the detection of element creations, we simplify the remaining constraints by referring to the creation timestamp of the new element instead of a global timepoint. For instance, in \autoref{fig:property}, the constraints of \emph{reac} are defined in relation to the \emph{cts} of  \emph{datum}. Finally, satisfaction of \emph{sat}($\kappa$) is decided by the OCL interpreter in EMF.

For step $\sigma_3$, we initially used the native EMF method to remove elements from the model based on the derived pruning rules. However, preliminary tests showed that the removal of elements incurred a significant increase in time. We discovered that due to the implementation of the references in EMF as array lists, the removal of an element causes all the other elements of the list to shift, making the time complexity of removing an element linear in the size of the list. In order to render the removal more scalable, we transparently replace the native EMF method with an optimized data structure and removal method via a Java agent.
\vspace{-1mm}
\subsection{Experimental Setting}

We evaluate \APPROACHNAME using a simulator for the SHS.\footnote{All experiments and simulations have been conducted on a Xeon E5-2643 and an OpenJDK8 JVM. Memory measurements are based on values reported by the JVM.}
We equip our simulator with a \mbox{MAPE-K} feedback loop that uses an architectural runtime model of the simulated SHS. In this paper we focus on the analysis phase of the adaptation loop and thus on the detection of missed antibiotic administrations, that is, $\mathcal{C}$ from the medical guideline in \cite{Wilson_2007_PreventionofInfectiveEndocarditisAGuidelineFromtheAmericanHeartAssociation}. 
We consider two implementation variants of \APPROACHNAME: \textbf{\VARIANTNAME}, where steps $\sigma_1$, $\sigma_2$ are implemented, and \textbf{\VARIANTNAMEplus}, where steps $\sigma_1$ and $\sigma_2$ plus the pruning of the model, i.e., step $\sigma_3$, are implemented. Only one variant is employed per simulation.

The interval of the \mbox{MAPE-K} loop is set to 3600 (an hour in seconds). 
Our simulator generates a random but repeatable sequence of timestamps picked from a range from zero to the number of seconds in a month. Each timestamp is associated to a number of change events---their number and type are parameterized and may vary. The simulation loop processes the event sequence and applies the associated changes on the runtime model. Each event corresponds to the creation of a node and the associated edge in the model, e.g., an event about a new datum corresponds to the creation of a \emph{datum} node and an edge to the emitting sensor. The simulation starts with an initial model size of 10 \emph{PatientSensors}, each connected to a \emph{Pump} via a \emph{Connector}---no other sensors or pumps are created. Then, at an almost constant rate, approx. the same amount of elements is created within each \mbox{MAPE-K} loop, that is, per loop, 1100 elements. 

In total, per \emph{PatientSensor}, 10000 events are generated which create a \emph{datum} with a value that is with an even chance either set to \emph{op} or \emph{noise} (an irrelevant \emph{datum}). Per \emph{Pump}, 30000 events are generated about reactions. At the end of the simulation, 400000 events have been processed which corresponds to approx. 800000 element additions to the model. Following each adaptation loop, a)  pruning is performed (only in \VARIANTNAMEplus) and b) the employed variant detects adaptation issues caused by events---by means of searches for matches of the condition of $\mathcal{C}$ in the runtime model.

Setting the rate of generated elements allows us to systematically increase the search space and thus investigate the scalability of \APPROACHNAME at detecting adaptation issues with respect to memory consumption and execution time of the pattern matching.

\subsection{Results}
\label{subsec:results}
\autoref{fig:evaluation-results} shows the execution times and memory consumption of the two variants for our simulation. In both figures, the x axis indicates the simulation timestamp. The number of processed events grows with the timestamp value. Each plot point represents the end of a loop and the execution of the query and, if applicable, pruning.

\begingroup
\setlength{\textheight}{\the\dimexpr19.3cm+1\baselineskip}
\setlength{\textfloatsep}{.7\baselineskip plus 0.2\baselineskip minus 0.5\baselineskip}
\setlength{\belowcaptionskip}{0pt}
\setlength{\abovecaptionskip}{0pt}
\begin{figure}
	\centering
	\begin{subfigure}[b]{.49\columnwidth}
		\centering
		\includegraphics[width=0.99\linewidth]{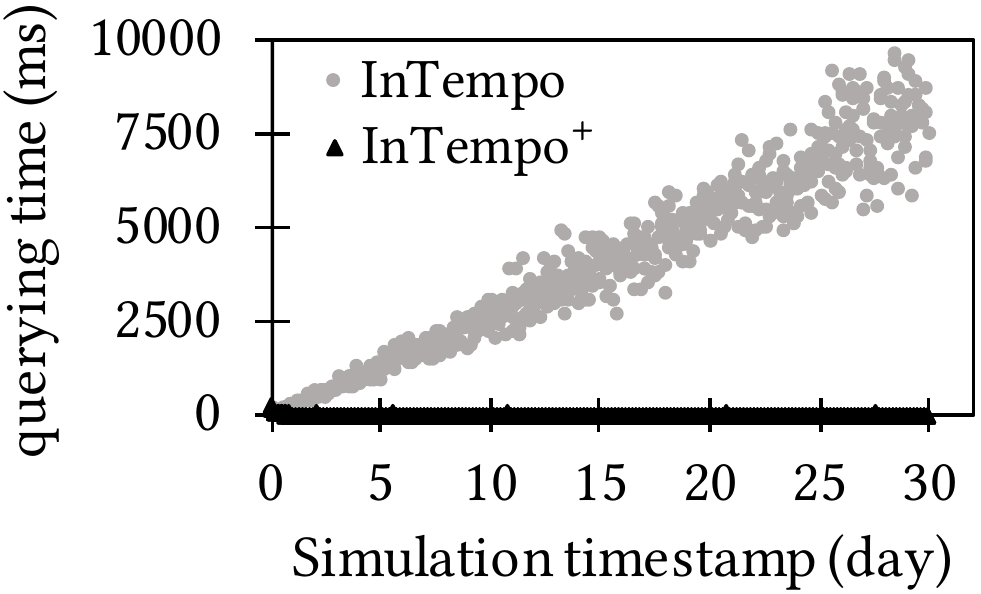}
	\end{subfigure}
	\begin{subfigure}[b]{.494\columnwidth}
		\centering
		\includegraphics[width=0.99\linewidth]{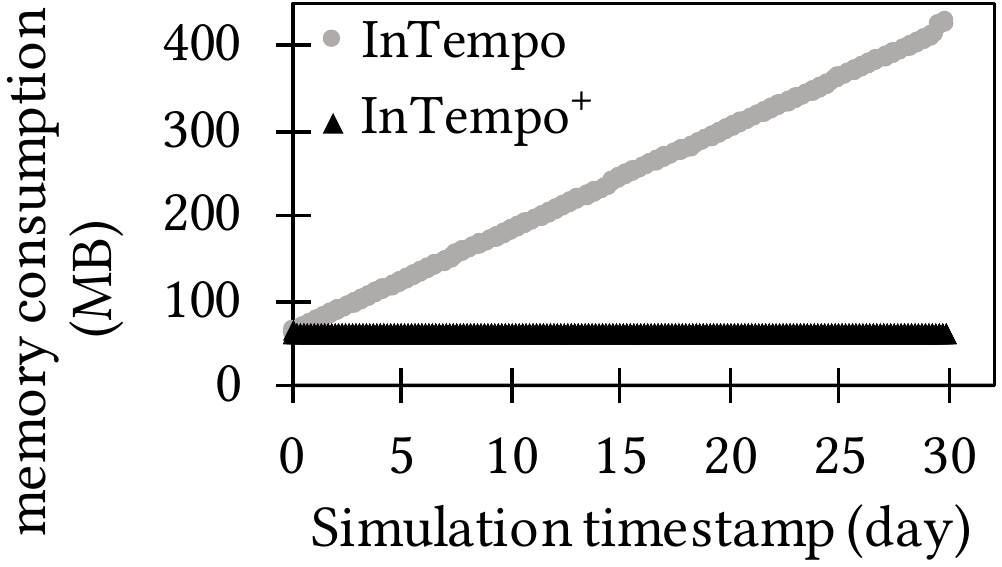}
	\end{subfigure}
	\caption{\label{fig:evaluation-results} Querying Time~(left) and Memory Consumption.}
	\vspace{-5mm}
\end{figure}
\endgroup

The plots of the \VARIANTNAME variant show that the execution time of the implementation grows linearly with respect to the size of the search space; as the number of \emph{reactions} grow, the number of potential matches increases as well. A similar effect is observed for the memory consumption of the \VARIANTNAME  variant, which strongly motivates the consideration of pruning in settings where the incurred increase of adaptation time---due to the duration of pruning---is acceptable. Regarding the \VARIANTNAMEplus variant, given the extensive pruning performed, the results are also to be expected; the pruning of all \emph{datum} nodes and the majority of \emph{reaction} nodes shows that the model size, the execution time, and the memory consumption of the variant are almost constant. This observation indicates that a variant with effective pruning rules can lead to a highly scalable runtime model that facilitates efficient pattern matching with temporal conditions. The peak in the execution time and memory consumption at the beginning of the simulation is due to implementation details: First, the ``warm-up'' rounds of the SD tool where necessary structures are initialized; Secondly, a starting graph is created at the beginning, before the \emph{datum} and \emph{reaction} nodes.

Our experiments show that the required time for searching and removing the elements is always less than 20ms i.e.,~20\% of the pattern execution time in each experiment. The reasons behind this result is the high pruning frequency, the simplicity of the pruning rules, and the optimized data structure and removal method presented in \autoref{subsec:implementation}. Pruning time is influenced by these factors and might  increase in different settings. Nevertheless we expect that applying pruning in executions with numerous events, provided an increase in the overall duration of the adaptation is allowed, will normally have a positive effect on the long-term scalability of the model. Furthermore, similarly to our  conducted experiments,  pruning might reduce the search space of the pattern matching, therefore ultimately decreasing the execution time of the matching. In our experiment, where we prune extensively, this effect is significant. In other settings, pruning might not result in such a considerable reduction of the search space.
	\vspace{-1mm}
\subsection{Threats to Validity}
Threats to internal validity concerns how we performed the experiments and interpreted the results. We systematically investigated the scalability and memory consumption of the alternative solutions 
by using a controlled simulation environment of a SHS. 
Particularly, we are interested in the effects of incremental checking and data pruning on the performance scalability and memory consumption of a pattern matching mechanism. To solely focus on these effects, the two compared variants share identical monitoring, planning, and execution phases of the MAPE-K feedback loop and they use the same architectural runtime model. Moreover, the experiments draw from a medical guideline where instructions are deterministic and typically of the form \emph{condition-action}, and as such, enable replicating a simulation for the two variants and over
multiple runs. This enables a fair comparison and the consideration of variations of the measured execution time for the pattern-matching. 

Threats to external validity may restrict the generalization of our evaluation results outside
the scope of our experiments.
We evaluated our approach on a simulated SHS using a synthesized event sequence. 
The systematic increase of the size of the sequence and, thus, the model size (number of patient sensors and pumps) allows for investigating the behavior of \APPROACHNAME under increasing complexity. While this investigation can serve as an indication, 
quantitative claims on scalability require extensive simulation scenarios that support a realistic setting for a large scale IoT system.
Synthetic data can limit the generality of experimental results. We nevertheless opted for it for a number of reasons: First, our focus was scalability whereas real sequences are typically used to assess fault-detection capabilities \cite{Dou_2017_AModelDrivenApproachtoTraceCheckingofPatternBasedTemporalProperties}. Secondly, because parameterizing data factors in a systematic way, such as the rate of generated events, helped us isolate and demonstrate key aspects of scalability that are relevant in IoT systems that produce voluminous streams of data. Finally, synthesized data allowed us to devise a scheme to monitor a real-world medical guideline which renders our approach relevant to IoT systems---particularly those related to healthcare.

\section{Related Work}\label{sec:related}
Capturing the history of a system execution either in a runtime model or a temporal graph database has been researched extensively (cf. \cite{Hartmann.2014.Native, Haeusler.2018.Combining, Brand.2019.Modeling}). Nonetheless, the scalable retrieval of information and the sustainability of the model has not received equal attention \cite{DBLP:journals/sosym/BencomoGS19}. The authors in \cite{Bencomo2019} build on a temporal graph database to store versions of a runtime model which is queried post-mortem by means of an OCL extension that supports temporal statements. We opt for an in-memory representation of our model, as shared memory space generally makes the real-time information retrieval faster, which is key for the systems of interest in this work. Additionally, model fragments evolve at different pace and thus versioning proves cumbersome on model level \cite{Hartmann.2014.Native}. Finally, implementation of a solution for the memory accumulation is missing from~\cite{Bencomo2019}. 

Event-based querying schemes and tools which rely on the (incremental) maintenance of a database have also been researched extensively. Here we discuss only those who support at least two of the key features of \APPROACHNAME: events containing data, temporal requirements, and graph-based representation of system entities and their relationships. The Metric First-order Temporal Logic \cite{DBLP:journals/jacm/BasinKMZ15} is as expressive as \MTGL and its supporting tool, MonPoly, could potentially use relations to represent runtime models and incrementally check formulas. However, the use of relations results in overly technical and error-prone system representations and formulation of pattern-based conditions \cite{Dou_2017_AModelDrivenApproachtoTraceCheckingofPatternBasedTemporalProperties}. Moreover, the tool performs inefficiently with past temporal conditions that require the storing of large amounts of data \cite{DBLP:conf/fmcad/HavelundPU17}.
An approach based on a runtime model and  graph structures being checked incrementally is presented in \cite{Bur_2018_DistributedGraphQueriesforRuntimeMonitoringofCyberPhysicalSystems,Bur_2019_Distributedgraphqueriesovermodelsruntimeforruntimemonitoringofcyberphysicalsystems} but does not support temporal requirements. 
The work in \cite{Dou_2017_AModelDrivenApproachtoTraceCheckingofPatternBasedTemporalProperties} processes propositional events that contain no data, makes changes on a model and then checks the satisfaction of OCL constraints which have been translated from a pattern-based specification.

In \cite{Giese_2019_MetricTemporalGraphLogicoverTypedAttributedGraphs} we have defined a formal operation for the reduction of an \MTGL condition with future operators to an NGC which can be checked against a graph that captures history with timestamps. However, the approach in  \cite{Giese_2019_MetricTemporalGraphLogicoverTypedAttributedGraphs} does not consider incremental checking nor a means to limit data accumulation.

\section{Conclusion and Future Work}\label{sec:conclusion}
%
%
Benefiting from model-driven engineering techniques, this paper proposes a memory-efficient scheme to incrementally check for conditions of adaptation rules which contain temporal requirements. The scheme relies on a runtime model with history that captures the evolution history of the system.
We efficiently prune the runtime model to only retain the information that are temporally relevant to the queries. This reduces the perpetual accumulation of historical information to a required minimum. As a result, the proposed approach is scalable and memory-efficient despite the growing number of change events. We confirmed the scalability and efficient memory consumption of the approach via empirical experiments on a simulated smart healthcare system. 

%
In the future, we plan to extend our work by formalizing the approach, evaluating it for other domains, supporting more complex conditions, automating the implementation, and finally developing our pruning strategy based on practical and theoretical considerations, e.g. pruning frequency and minimum overhead respectively.

\bibliographystyle{ACM-Reference-Format}
\balance\bibliography{biblio/seams20}

\end{document}